\documentclass{article}
\usepackage[preprint, nonatbib]{nips_2018}
\usepackage[square,numbers]{natbib}
\usepackage[utf8]{inputenc} 
\usepackage[T1]{fontenc}    
\usepackage{hyperref}       
\usepackage{url}            
\usepackage{booktabs}       
\usepackage{nicefrac}       
\usepackage{microtype}      
\usepackage{graphicx}
\usepackage{tabularx}
\usepackage{hyperref}

\usepackage{amsmath,amssymb,amsfonts}
\usepackage{algorithmic}
\usepackage{textcomp}
\usepackage{epstopdf}
\usepackage{float}
\usepackage{array}
\usepackage{multirow}
\usepackage{bm}
\usepackage{stfloats}

\title{A Lightweight Structure Aimed to Utilize Spatial Correlation for Sparse-View CT Reconstruction}
\author{
  Yitong Liu \\
  \texttt{liuyitong@bupt.edu.cn} \\
\And
 Ken Deng \\
  \texttt{arieldeng@bupt.edu.cn} \\
\And
Chang Sun \\
  \texttt{sc1998@bupt.edu.cn} \\
  \And
Hongwen Yang \\
  \texttt{yanghong@bupt.edu.cn} \\
}

\begin{document}
\maketitle
\vspace{-1em}

\begin{abstract}
Sparse-view computed tomography (CT) is known as a widely used approach to reduce radiation dose while accelerating imaging through lowered projection views and correlated calculations. However, its severe imaging noise and streaking artifacts turn out to be a major issue in the low dose protocol. In this paper, we propose a dual-domain deep learning-based method that breaks through the limitations of currently prevailing algorithms that merely process single image slices. Since the scanned object usually contains a high degree of spatial continuity, the obtained consecutive imaging slices embody rich information that is largely unexplored. Therefore, we establish a cascade model named LS-AAE which aims to tackle the above problem. In addition, in order to adapt to the social trend of lightweight medical care, our model adopts the inverted residual with linear bottleneck in the module design to make it mobile and lightweight (reduce model parameters to one-eighth of its original) without sacrificing its performance. In our experiments, sparse sampling is conducted at intervals of 4°, 8° and 16°, which appears to be a challenging sparsity that few scholars have attempted. Nevertheless, our method still exhibits its robustness and achieves the state-of-the-art performance by reaching the PSNR of 40.305 and the SSIM of 0.948, while ensuring high model mobility. Particularly, it still exceeds other current methods when the sampling rate is one-fourth of them, thereby demonstrating its remarkable superiority.
\end{abstract}
\vspace{-1em}

\section{Introduction}
Over the last few decades ~\cite{Cormack1964Representation, Hounsfield1974Computerized}, X-ray Computed Tomography (CT) has demonstrated its prominent practical value and wide range of applications including clinical diagnosis, safety inspection and industrial detection ~\cite{Wang2008An}. Especially in the past year, due to the global spread of the Corona Virus Disease 2019 (COVID-19), the term CT has become well-known to the public as an essential auxiliary technology. However, the radiation dose brought by CT has a nonnegligible side effect on the human body. Since it has a latent risk of inducing cancers, radiation dose reduction is becoming more and more crucial under the principle of ALARA (as low as reasonably achievable) ~\cite{Krishnamoorthi2011Effectiveness, Slovis2002CT, Mccollough2009Strategies, Mccollough2006CT}.

Generally speaking, there are two approaches to reduce radiation dose. The approach of tube current (or voltage) reduction ~\cite{Poletti2007Low-DoseVersus, Tack2003DoseReduction} lowers the x-ray exposure in each view but suffers from the increased noise in projections. Although the approach of projection number reduction ~\cite{Bian2010Evaluation, Bian2012Optimization} (also known as sparse-view CT) can avoid the former problem and realize the additional benefit of accelerated scan and calculation, it leads to severe image quality degradation of increased streaking artifacts brought by its missing projections. In this paper, we focus on effectively repairing and reconstructing sparse-view CT so as to acquire high-quality CT images.

Sparse-view CT reconstruction has always been a classic inverse problem which has attracted wide attention \cite{Jin2016Deep}. In the past few decades, iterative reconstruction methods have become the dominant approach to solve inverse problems \cite{Andersen1984Simultaneous, Wu2017Iterative, Hu2017An, Zhang2019JSR}. With the advent of compressed sensing \cite{Candes2006Robust} and its related regularizers, the quality of reconstructed images has been improved to a certain extent. One of the most typical regularizers is the total variation (TV) method, algorithms based on which include TV-POCS \cite{Sidky2008Image}, TGV method \cite{Niu2014Sparse}, SART \cite{Andersen1984Simultaneous} and SART-TV \cite{Sidky2009Accurate} etc. In addition, dictionary learning is also commonly used as a regularizer. For example, \cite{Xu2012Low} constructs a global dictionary and an iterative adaptive dictionary to solve the problem of low-dose CT reconstruction.

In recent years, with the improvement of computing power, there comes a rapid growth in deep learning \cite{LeCun2015DeepLearning}. Subsequently, neural networks have been widely applied in image analysis tasks, such as image classification \cite{Russakovsky2015ImageNet}, image segmentation \cite{Long2015Fully, Ronneberger2015U, Soltaninejad2020Three}, especially inverse problems in image reconstruction, such as artifacts reduction \cite{Dong2015Compression, Guo2016Building}, denoising \cite{Xie2012Image} and inpainting \cite{Kulkarni2016ReconNet}. Since GAN (Generative Adversarial Networks) was designed elaborately by Goodfellow in 2014 \cite{goodfellow2014generative}, it has been adopted in many image processing tasks due to its prominent performance in realistically predicting image details. Therefore, GANs are also naturally applied to improving the quality of low-dose CT images \cite{Bai2018Limited, Xie2019Artifact, Zhao2018Sparse}. In addition, Ye et al. explored the relationship between deep learning and classical signal processing methods in \cite{Ye2017Deep}, explained the reason why deep learning can be employed in imaging inverse problems, and provided a theoretical basis for the application of deep learning in low-dose CT reconstruction.

Some researchers adopt deep learning-based architectures to complement and restore the limit-view Radon data \cite{Dong2019A, fu2020a, Bai2018Limited, Anirudh2018Lose, Dai2019Limited, ghani2018deep}. Dong et al. \cite{Dong2019A} used U-Net \cite{Ronneberger2015U} to predict the missing Radon data, then reconstruct it to the image through FBP \cite{katsevich2002theoretically}. Jian Fu et al. \cite{fu2020a} built a network that involves the tight coupling of the deep learning neural network and DPC-CT (Differential phase-contrast CT) reconstruction algorithm in the domain of DPC projection sinograms. The estimated result is a complete phase-contrast projection sinogram. Rushil Anirudh et al. established CTNet \cite{Anirudh2018Lose}, a system of 1D and 2D convolutional neural networks, which operates on the limited-view sinogram to predict the full-view sinogram, and then fed it to the standard analytical and iterative reconstruction algorithms to obtain the final result.

Other researchers carried out post-processing on reconstructed images with deep learning models, so as to remove the artifacts and noises for upgrading the quality of these images\cite{Han2018Framing, Zhang2018A, zhang2016image, Xie2019Artifact, Wang2020Deep, kuanar2019low, guan2020fully}. In 2016, a deep convolutional neural network \cite{zhang2016image} was proposed to learn an end-to-end mapping between the FBP and artifact-free images. In 2018, Yoseob Han and Jong Chul Ye designed a dual frame and tight frame U-Net \cite{Han2018Framing} which satisfies the frame condition and performs better for recovery of high frequency edges in sparse-view CT. In 2019, Xie et al. \cite{Xie2019Artifact} built an end-to-end cGAN model with joint function used for removing artifacts from limited-angle CT reconstruction images. In 2020, Wang et al. \cite{Wang2020Deep} developed a limited-angle TCT image reconstruction algorithm based on U-Net, which could suppress the artifacts and preserve the structures. Experiments have shown that U-Net-like structures are efficacious for image artifacts removal and texture restoration \cite{Ye2017Deep,Dong2019A,Han2018Framing,Wang2020Deep,guan2020fully} .

Since neural networks are capable of predicting unknown data in the Radon and image domains, a natural idea is to combine these two domains \cite{Lee2018High,liang2018comparison,Zhao2018Sparse,Zhu2020Low,hammernik2017a,Zhang2020Artifact} to acquire better restoration results. Specifically, it first complements the Radon data, and then remove the residual artifacts and noises on images converted from the full-view Radon data. In 2018, Zhao et al. proposed SVGAN \cite{Zhao2018Sparse}, an artifacts reduction method for low-dose and sparse-view CT via a single model trained by GAN. In 2019, Liang et al. \cite{liang2018comparison} proposed a comprehensive network combining projection and image domains. The projection estimation network is based on Res-CNN structure, and the image domain network takes the advantage of U-Net. In 2020, Zhu et al. designed ADAPTIVE-NET \cite{Zhu2020Low} to conduct joint denoising on the acquired sinogram and the reconstructed CT image, while reconstructing CT image in an end-to-end manner. In the past three years, experiments have proved that this sort of two-stage algorithm is quite conducive to image quality improvement.

All the current mainstream methods mentioned above make us notice that they solely process on each single CT image, while neglecting the solid fact the scanned object is always highly continuous. Consequently, there is abundant spatial information lies in these obtained consecutive CT images, which is largely left to be exploited. This enlightens us to propose a novel cascade model called LS-AAE (Lightweight Spatial Adversarial Autoencoder) that mainly focus on availably utilizing the spatial information between greatly correlated images. It has been proved in our experiments that this sort of structure design manages to efficaciously remove streaking artifacts in sparse-view CT images, and outruns other prevailing methods with its remarkable performance.

It is the social trend now to make healthcare mobile and portable. In lots of deep learning-based methods, however, scholars improve accuracy at the expense of sacrificing computing resources. Such computational complexity usually exceeds the capabilities of many mobile and embedded applications. This paper adopts the inverted residual with linear bottleneck \cite{Sandler2018MobileNetV2} in the module design to propose a mobile structure that reduce model parameters to one-eighth of its original without sacrificing its performance.

Although enhancing the sparsity of sparse-view CT can bring benefits of accelerated scanning and related calculations, it will cause additional imaging damage. Balancing image quality and X-ray dose level has become a well-known trade-off problem. Thus, in order to explore the limit of sparsity in sparse-view CT reconstruction, we conduct sparse sampling at intervals of 4°, 8° and most importantly, 16°. Even under such sampling sparsity, our model can still exhibit its remarkable robustness and the state-of-the-art performance.

We introduce our proposed method exhaustively in Section II, the experimental results and corresponding discussion are described in section III, and conclusion is stated in section IV.

\section{Methods}
\subsection{Preliminaries}
\subsubsection{Utilize Spatial Information for Artifact Removal}

As is known to all, consecutive CT images usually contain high spatial coherency since the scanned object is usually spatially continuous. On account of that, we can imagine these CT images as adjacent frames in a video which contains much more information than a still image. This high correlation within the sequence of images can improve the performance of artifact removal from two aspects. Firstly, the extension of search regions from two-dimensional image neighborhoods to three-dimensional spatial neighborhoods provide extra information which can be used to denoise the reference image. Secondly, using spatial neighbors helps to reduce streaking artifacts as the residual error in each image is correlated.

Also, we cannot help but notice that the task of artifact removal between consecutive images is similar to video denoising. Therefore, after investigating lots of research work on video denoising \cite{pascanu2013on, caballero2017real, maggioni2012video, arias2018video, vogels2018denoising, ehret2019model, claus2019videnn, davy2018non, tassano2019dvdnet, chen2016deep}, we find out that current state-of-the-art methods lay lots of emphasis on motion estimation due to the strong redundancy along motion trajectories. To conclude, in order to more effectively remove streaking artifacts from sparse-view CT images, we need to design a structure that can not only look into the three-dimensional spatial neighborhood, but also capture the motion between consecutive images.

\subsubsection{Enhance Mobility of Neural Networks}
In recent years, lots of research has been invested into tuning deep neural networks to achieve an optimal balance between efficiency and performance. Among them, depthwise separable convolutions \cite{howard2017mobilenets} exhibits its extraordinary capability and has gradually become an essential building block for numerous lightweight neural networks \cite{howard2017mobilenets, chollet2017xception, zhang2018shufflenet}. It aims to decompose the standard convolutional layer into two separate layers, namely the depthwise convolutional layer and the pointwise convolutional layer. The former layer is designed to perform lightweight filtering through employing a single convolutional filter per input channel, the latter one conducts $1\times1$ convolution to construct new features by computing linear combinations of input channels.

For the standard convolutional layer with input tensor size $(c_{in}, h, w)$, kernel size $(c_{out}, c_{in}, k, k)$ and output tensor size $(c_{out}, h, w)$, its computational cost equals to $c_{in} \cdot h \cdot w\cdot (k^2 \cdot  c_{out})$. However, in depthwise separable convolutions, the depthwise convolutional layer has a computational cost of $c_{in} \cdot  h \cdot w \cdot k^2$ since it merely operates on a single input channel, and the pointwise convolutional layer has a computational cost of $c_{in} \cdot  h \cdot  w \cdot  c_{out}$. Therefore, we only need a computational cost of $h \cdot  w \cdot  c_{in} \cdot  (k^2 + c_{out}) $ for depthwise separable convolutions, which is almost the one-ninth ($k$ equals to 3 in our case) of the standard convolution. Most importantly, depthwise separable convolutions manage to lower the computational complexity to a large extent without sacrificing its accuracy, which would make it perfect to be inserted into our module design.

\subsection{Overall Structure}
\subsubsection{Structure Overview}

\begin{figure*}[!htb]
\setlength{\abovecaptionskip}{-5pt}
\centerline
{\includegraphics[width=\linewidth]{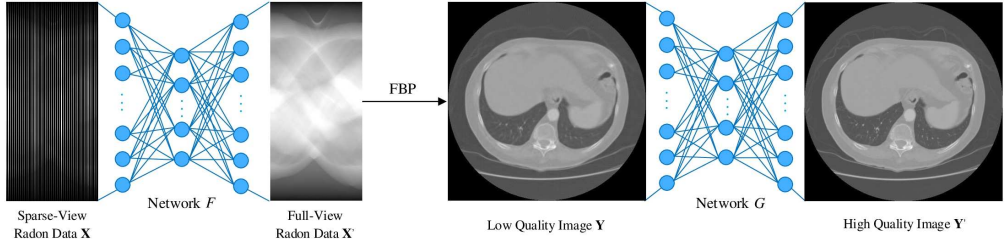}}
\caption{Structure overview. The sparse-view Radon data $\bold{X}$ is first sent to the neural network $F$ for completion, then the restored full-view Radon data $\bold{X}^\prime$ is converted to the image $\bold{Y}$, which is feed into the neural network $G$ for artifacts removal and we can finally obtain the ideal high-quality image $\bold{Y}^\prime$.}
\label{fig1}
\end{figure*}

We can learn from the universal approximation theorem \cite{hornik1989multilayer} that multilayer feedforward networks are capable of approximating various continuous functions. This inspires us to think that neural networks can be used to learn complex mappings that are difficult to solve through mathematical analysis. Thus, in this paper, we utilize a deep learning-based structure that combines the Radon domain and the image domain (Figure \ref{fig1}) to solve the task of sparse-view CT reconstruction and inpainting.

Firstly, we want to make full use of the prior information in the Radon domain by converting the sparse-view Radon data $\bold{X}$ to the full-view Radon data $\bold{X}^{\prime}$ so as to complement the missing data in some scanning angles. This process can be represented by the mapping: $\bold{X} \xrightarrow{f} \bold{X}^{\prime}$ according to the universal approximation theorem, where function $f$ can be approximated through our proposed neural network $F$. After we obtain the full-view Radon data $\bold{X}^{\prime}$, we transform it to the image $\bold{Y}$ through FBP. Although the first stage manages to alleviate the severe imaging damage from the original sparse-view CT image, there are still lots of streaking artifacts existing in Y that need to be removed to acquire the high-quality restored result $\bold{Y}^{\prime}$. We represent the restoration process into the mapping: $\bold{Y} \xrightarrow{g} \bold{Y}^{\prime}$, where function $g$ can be approximated through our proposed neural network $G$. Through the above two-stage structure that combines the Radon domain with the image domain, we can finally get the ideal restored results.

\subsubsection{Stage One: Data Completion in the Radon Domain}
\begin{figure}[!tb]
\vspace{-7pt}
\setlength{\abovecaptionskip}{-5pt}
\centerline{\includegraphics[scale=0.7]{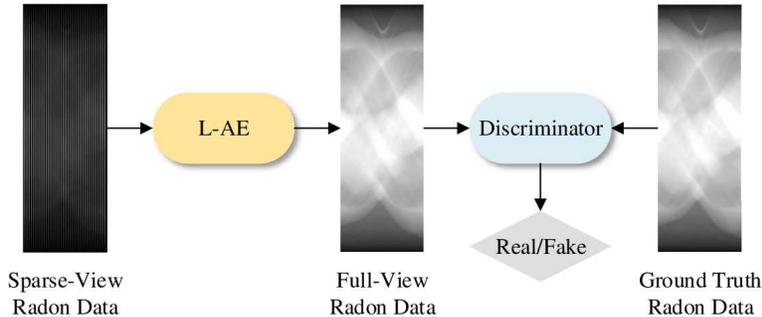}}
\caption{The diagram of our proposed L-AAE, which is composed of a L-AE and a discriminator that help restore the image texture.}
\label{fig2}
\vspace{-5pt}
\end{figure}

\begin{figure}[!tb]
\setlength{\abovecaptionskip}{-10pt}
\centerline{\includegraphics[width=\linewidth]{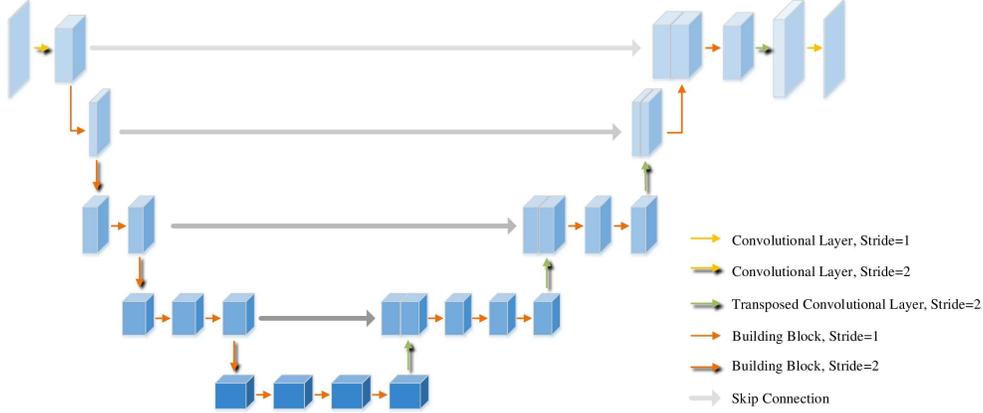}}
\caption{The detailed structure of the L-AE. Input images are first feed into the encoder for feature extraction and then sent into the decoder for texture restoration, where skip connections are added to merge low-level features.}
\label{fig3}
\end{figure}

We first adopt linear interpolation to convert the original sparse-view Radon data to full-view Radon data so as to satisfy the structural characteristics of our proposed neural network, which requires the input and output images to have the same resolution. Then we build a lightweight adversarial autoencoder (L-AAE) in Figure \ref{fig2} to restore the Radon data, the structure of its autoencoder (L-AE) can be seen from Figure \ref{fig3} and Table \ref{table1}, which is composed of the encoder and the decoder that are highly symmetrical.

\begin{table}[!htb]
  \centering
  \caption{Parametric structure of the L-AE}
  \renewcommand{\arraystretch}{1.3}
    \begin{tabular}{m{5.5em}<{\centering} m{5.5em}<{\centering}cc m{4em}<{\centering} m{4em}<{\centering} } \hline \hline
    \multicolumn{1}{m{5.5em}<{\centering}}{Layer} & \multicolumn{1}{m{5.5em}<{\centering} }{$IC$} & \multicolumn{1}{m{1.6em}<{\centering} }{$OC$} & \multicolumn{1}{m{2em} <{\centering}}{Stride} & Input Size & Output Size\\\hline
    Conv1 & 1     & 32    & 2     & 192$\times$512 & 96$\times$256 \\
    Block1 & 32    & 16    & 1     & 96$\times$256 & 96$\times$256 \\
    Block2\_1 & 16    & 32    & 2     & 96$\times$256 & 48$\times$128 \\
    Block2\_2 & 32    & 32    & 1     & 48$\times$128 & 48$\times$128 \\
    Block3\_1 & 32    & 64    & 2     & 48$\times$128 & 24$\times$64 \\
    Block3\_2 & 64    & 64    & 1     & 24$\times$64 & 24$\times$64 \\
    Block3\_3 & 64    & 64    & 1     & 24$\times$64 & 24$\times$64 \\
    Block4\_1 & 64    & 128   & 2     & 24$\times$64 & 12$\times$32 \\
    Block4\_2 & 128   & 128   & 1     & 12$\times$32 & 12$\times$32 \\
    Block4\_3 & 128   & 128   & 1     & 12$\times$32 & 12$\times$32 \\
    Block4\_4 & 128   & 128   & 1     & 12$\times$32 & 12$\times$32 \\
    Trans Conv1 & 128   & 64    & 2     & 12$\times$32 & 24$\times$64 \\
    Block5\_1 & \multicolumn{1}{m{6em}}{64+64(Concat)} & 64    & 2     & 24$\times$64 & 24$\times$64 \\
    Block5\_2 & 64    & 64    & 1     & 24$\times$64 & 24$\times$64 \\
    Block5\_3 & 64    & 64    & 1     & 24$\times$64 & 24$\times$64 \\
    Trans Conv2 & 64    & 32    & 2     & 24$\times$64 & 48$\times$128 \\
    Block6\_1 & \multicolumn{1}{m{6em}}{32+32(Concat)} & 32    & 1     & 48$\times$128 & 48$\times$128 \\
    Block6\_2 & 32    & 32    & 1     & 48$\times$128 & 48$\times$128 \\
    Trans Conv3 & 32    & 16    & 2     & 48$\times$128 & 96$\times$256 \\
    Block7 & \multicolumn{1}{m{6em}}{16+16(Concat)} & 32    & 1     & 96$\times$256 & 96$\times$256 \\
    Block8 & \multicolumn{1}{m{6em}}{32+32(Concat)} & 32    & 1     & 96$\times$256 & 96$\times$256 \\
    Trans Conv4 & 32    & 16    & 2     & 96$\times$256 & 192$\times$512 \\
    Conv9 & 16    & 1     & 1     & 192$\times$512 & 192$\times$512 \\ \hline \hline
    \end{tabular}%
  \label{table1}%
  \vspace{-10pt}
\end{table}%

We perform four down sampling in the encoder to obtain high-level semantic features of the input image, which is initially downsampled through conv1 layer with a stride of 2, and the subsequent downsampling is separately accomplished by the first building block of each unit. Each downsampling will halve the height and width of the activation map and double the number of channels.

As for the decoder, we correspondingly conduct four upsampling to restore the texture of the input image. Deconvolution is adopted here for upsampling with its kernel size and stride both equal to 2, so that each upsampling will double the height and width of the activation map and halve the number of channels. In addition, we add skip connections \cite{he2016identity} between the encoder and decoder feature maps of the same resolution. Since the final feature map of the encoder has a relatively low resolution due to multiple downsampling, it will have an undesirable effect on the restoration of the image texture in the decoder. While the skip connection incorporates low-level features from the encoder which have a high resolution and contain abundant detailed information that will help to accurately restore the image texture. This sort of multi-scale, U-Net-like architectures have been proved to be effective in processing medical images.

\begin{figure*}[!htb]
\setlength{\abovecaptionskip}{3pt}
\centerline{\includegraphics[width=14cm]{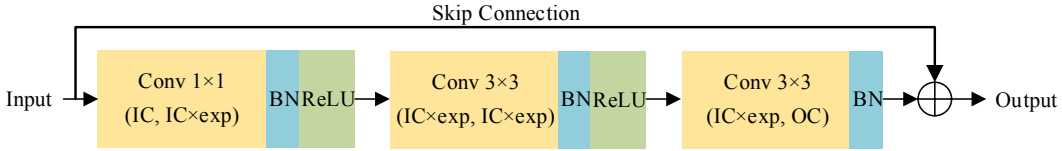}}
\caption{The detailed diagram of the building blocks in L-AAE, which adopt the inverted residual with linear bottleneck.}
\label{fig4}
\end{figure*}

The detailed structure of our building block can be seen from Figure \ref{fig4}, it adopts the inverted residual with linear bottleneck referring to \cite{Sandler2018MobileNetV2}, each block is composed of three convolutional layers. The first layer expands (characterized by the expansion factor $exp$) a low-dimensional compressed representation to high dimension with a kernel size of $1 \times 1$. The intermediate expansion layer adopts lightweight depthwise convolutions mentioned above so as to significantly decreases the number of operations (ops) and memory needed while sustaining the same performance. The last layer projects the feature back to a low-dimensional representation with a linear convolution like the first layer. All these layers are followed by batch normalization \cite{ioffe2015batch} and ReLU \cite{glorot2011deep} as the non-linear activation except for the last layer that only followed by a batch normalization layer.

In Figure \ref{fig4}, $IC$ and $OC$ stand for the input and output channel of building blocks respectively. All convolutional layers in all building blocks have a stride of 1 except for Block2\_1, Block3\_1 and Block4\_1 that have a stride of 2 to conduct downsampling.Expansion factor $exp$ is 1 for Block1, Block7 and Block8 to avoid large ops and memory cost, we set up $exp$ to be 3 for Block5\_1 and Block6\_1, and every block expect these mentioned above have an $exp$ of 6. Besides, shortcut connections are implemented in blocks that have the same resolution between its input and output feature maps to enhance information flow and also improve the ability of a gradient to propagate across multiplier layers. We adopt $1\times1$ convolution in shortcuts when the number of channel in the input and output feature maps is different.

The discriminator in our L-AAE aims to strengthen model's ability to restore the detailed texture of images, its structure is almost the same as the encoder above, except that its Block4\_3 and Block4\_4 have an $OC$ of 64 and 1 respectively. The output of Block4\_4 is flattened and sent to sigmoid function for probability prediction, which we average to get the final output that represents the input image's probability to be a real image. This novel lightweight AAE enables us to acquire the well restored Radon data that are complete in every scanning angle, and the computational cost is about 8 times smaller than that of standard convolutions without sacrificing its accuracy.

\subsubsection{Stage Two: LS-AAE -- Image Inpainting through spatial information}
After stage one, we transform the acquired full-view Radon data to images and find out that, we successfully enrich the information in the Radon domain and alleviate streaking artifacts from the original sparse-view CT imaging. Now in stage two, we will mainly focus on removing artifacts, restoring image to an ideal level. As mentioned above, we need a neural network that not only look into the three-dimensional spatial neighborhood, but also capture the motion between consecutive images, so as to efficaciously utilize the abundant spatial information between consecutive images to remove artifacts from the input image.

\begin{figure*}[!htb]
\setlength{\abovecaptionskip}{-10pt}
\centerline{\includegraphics[width=\linewidth,scale=0.8]{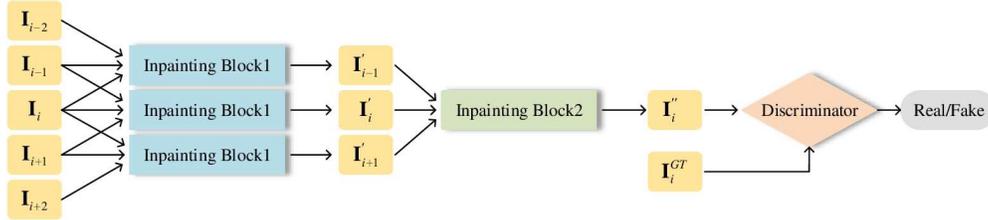}}
\caption{The diagram of our proposed LS-AAE. It combines an autoencoder that fully utilizes the spatial correlation between consecutive CT images and a discriminator that help refine image details.}
\label{fig5}
\end{figure*}

Generally speaking, motion estimation always brings an additional degree of complexity that is adverse to model's implementation in reality. It means that we need a structure that can manage to deploy motion estimation without much resource cost, we refer to \cite{tassano2020fastdvdnet} and its general structure appears to be a cascaded two-step architecture that inherently embed the motion of objects. Inspired by this, we propose a model named Lightweight Spatial Adversarial Autoencoder (LS-AAE) which can be seen from Figure \ref{fig5}. It slightly modifies the L-AE from Figure \ref{fig3} as its inpainting block, details are shown in Table \ref{table1}. The replacement from 2D convolution to 3D convolution enables our model to look into the three-dimensional spatial neighborhood for extra information.

\begin{table}[htbp]
  \centering
  \caption{From 2D convolution to 3D convolution}
  \renewcommand{\arraystretch}{1.3}
    \begin{tabular}{m{4em}<{\centering}m{3.5em}<{\centering}m{1.4em}<{\centering}m{1.4em}<{\centering}m{2.5em}<{\centering}m{2.5em}<{\centering}m{2.8em}<{\centering}}\hline\hline
          & Layer & \multicolumn{1}{m{1.4em}<{\centering}}{$IC$} & \multicolumn{1}{m{1.4em}<{\centering}}{$OC$} & Kernel Size & Stride & Padding \\ \hline
    \multicolumn{1}{c}{2D Convolution} & Conv1 & 1     & 16    & (3,3) & (2,2) & (1,1) \\ \hline
    \multicolumn{1}{c}{\multirow{2}[4]{*}{3D Convolution}} & Conv1\_1 & 1     & 16    & (3,3,3) & (1,2,2) & (1,0,0) \\
     & Conv1\_2 & 16    & 32    & (3,3,3) & (2,1,1) & (0,1,1) \\ \hline\hline
    \end{tabular}%
  \label{table2}%
\end{table}%

As shown in Figure \ref{fig5}, five consecutive images $\left\{\bold{I}_{i-2}, \bold{I}_{i-1}, \bold{I}_i, \bold{I}_{i+1}, \bold{I}_{i+2}\right\}$ are sent into the LS-AAE to restore the middle one. We firstly treat these inputs as triplets of consecutive images $\left\{\bold{I}_{i-2}, \bold{I}_{i-1}, \bold{I}_i\right\}$, $\left\{\bold{I}_{i-1}, \bold{I}_i, \bold{I}_{i+1}\right\}$ and $\left\{\bold{I}_i, \bold{I}_{i+1}, \bold{I}_{i+2}\right\}$, then enter them into the Inpainting Blocks 1. Subsequently, we obtain the outputs of these blocks and combine them into triplet $\left\{\bold{I}^{\prime}_{i-1}, \bold{I}^{\prime}_i, \bold{I}^{\prime}_{i+1}\right\}$ which will be sent into Inpainting Block 2 to acquire the ultimate estimation $\bold{I}^{\prime\prime}_i$ corresponding to the central image $\bold{I}_i$. The LS-AAE digs deep into the three-dimensional space and implicitly handles motion without any explicit motion compensation stage on account of the traits of its architecture. Besides, the three Inpainting Blocks in step one share the same weights so as to avoid memory cost. We also add a discriminator in stage two to better restore the image texture, the predicted image $\bold{I}^{\prime\prime}_i$ and its corresponding ground truth (the full-view CT imaging) $\bold{I}^{GT}_i$ are both send into this discriminator, its structure is exactly the same as it is in stage one..

\subsection{Network Training}
Stage one and stage two are trained separately. For the autoencoders in these two models, we employ the multi-loss function below, which is consists of three parts $l_{MSE}$, $l_{Adv}$ and $l_{Reg}$ with their respective hyperparameters $\alpha_1$, $\alpha_2$ and $\alpha_3$.

\begin{equation}
\label{eq1}
{{l}_{AE}}={{\alpha}_{1}}{{l}_{MSE}}+{{\alpha}_{2}}{{l}_{Adv}}+{{\alpha}_{3}}{{l}_{Reg}}
\end{equation}

$l_{MSE}$ calculates the mean square error between the restored image and its corresponding ground truth, it is widely used in various image inpainting tasks since it provides an intuitive evaluation for the model's prediction. The expression of  $l_{MSE}$ can be seen from Equation (\ref{eq2}).

\begin{equation}\label{eq2}
{{l}_{MSE}}=\frac{1}{W\times H}\sum\limits_{x=1}^{W}{\sum\limits_{y=1}^{H}{{{(\bold{I}_{x,y}^{GT}-{{G}_{AE}}{{({{\bold{I}}^{Input}})}_{x,y}})}^{2}}}}
\end{equation}

\noindent Where function $G_{AE}$ stands for the autoencoder, $\bold{I}^{Input}$ and $\bold{I}^{GT}$ are the input image and its corresponding ground truth, $W$ and $H$ are the width and height of the input image respectively.

$l_{Adv}$ refers to the adversarial loss. The autoencoder manages to fool the discriminator by making its prediction as close to the ground truth as possible, so as to achieve the ideal image restoration outcome. Its expression can be seen from Equation (\ref{eq3}).

\begin{equation}\label{eq3}
{{l}_{adv}}=1-D({{G}_{AE}}({\bold{I}^{Input}}))
\end{equation}

\noindent Where function $D$ and $G_{AE}$ stands for the discriminator and the autoencoder respectively, $\bold{I}^{Input}$ is the model's input image.

$l_{Reg}$ is the regularization term of our multi-loss function. Since noises will have a side effect on our restoration result, we add a regularization term to maintain the smoothness of the image and also prevent overfitting. TV Loss is widely used in image analysis tasks, it reduces the variation between adjacent pixels to a certain extent. Its expression can be seen from Equation (\ref{eq4}).

\begin{equation}\label{eq4}
{{l}_{Reg}}=\frac{1}{W\times H}\sum\limits_{x=1}^{W}{\sum\limits_{y=1}^{H}{\left\| \left. \nabla {{G}_{AE}}{{({\bold{I}^{Input}})}_{x,y}} \right\| \right.}}
\end{equation}

\noindent Where function $G_{AE}$ represents the autoencoder, $I^{Input}$ is the model's input image, $W$ and $H$ are the width and height of the input image respectively.  $\nabla$ calculates the gradient, $\left\| \cdot \right\|$ obtains the norm.

To optimize the discriminator of these two stages, their loss function should enable them to better distinguish between real and fake inputs. The loss function $l_{Dis}$ is shown in Equation (\ref{eq5}).

\begin{equation}\label{eq5}
{{l}_{D\text{is}}}=1-D\left( {\bold{I}^{GT}} \right)+D\left( {{G}_{AE}}\left( {\bold{I}^{Input}} \right) \right)
\end{equation}

\noindent Where function $D$ and $G$ stands for the discriminator and the autoencoder respectively, $\bold{I}^{Input}$ and $\bold{I}^{GT}$ are the input image and its corresponding ground truth. The discriminator outputs a scalar between 0 to 1 which represents the probability that the input image is real. Therefore, minimizing $1-D(\bold{I}^{GT})$/maximizing $D(\bold{I}^{GT})$ enables the discriminator to recognize real images, while minimizing $D(G_{AE}(\bold{I}^{Input}))$ enables the discriminator to distinguish fake images that generated from the autoencoder from all input images.

During the training process, we adopt the Adam algorithm \cite{kingma2015adam} for optimization. the learning rate is set to 1e-4 initially. For the multi-loss function, $\alpha_1$, $\alpha_2$ and $\alpha_3$ are set to 1, 1e-3, and 2e-8 respectively. We implement our whole structure using PyTorch \cite{paszke2017automatic} on two GeForce RTX 2080 Ti.

\section{Experiments}
We adopt the LIDC-IDRI \cite{armato2011the} as our dataset, which includes 1018 cases and approximately 240,000 DCM files of corresponding CT images. Cases 1 to 100 are divided into test set, cases 101 to 200 are divided into validation set, and the rest are divided into train set. Such a large amount of data allows us to train our models from scratch without overfitting. We utilize NumPy to read from these DCM files and conduct sparse sampling at intervals of 4, 8 and 16 (the corresponding full-view Radon data has 180 projections). Subsequently, we first analyze our overall structure through a series of ablation studies, and then compare our experimental results with other current methods to prove its superiority and robustness.

\subsection{Ablation Study}
With all these innovations we make in our overall structure design, it would be appropriate for us to conduct corresponding ablation studies to prove their necessity. In this part, all the experimental results are acquired from sparse-view CT data with an interval of 4 if there is no specific mention.

\subsubsection{The L-AE's Trade-off between Mobility and Performance}
As is known to all, U-Net has extraordinary performance in numerous medical image processing tasks, \cite{Han2018Framing} implemented it for sparse-view CT imaging restoration and obtained outstanding restoration results. To testify that our proposed autoencoder can achieve a good balance between performance and mobility, we replace it with U-Net in the first stage and compare the restoration results and model parameters of this stage with ours, as shown in Table \ref{table3}. The images mentioned in Table \ref{table3} are reconstructed from the Radon data restored through stage one.

\begin{table}[htbp]
  \centering
  \caption{U-Net VS. L-AE}
   \renewcommand{\arraystretch}{1.3}
    \begin{tabular}{m{3em}<{\centering}ccccm{4.945em}<{\centering}}\hline\hline
    \multicolumn{1}{c}{} & \multicolumn{1}{m{3em}<{\centering}}{Radon PSNR} & \multicolumn{1}{m{3em}<{\centering}}{Radon SSIM} & \multicolumn{1}{m{3em}<{\centering}}{Image PSNR} & \multicolumn{1}{m{3em}<{\centering}}{Image SSIM} & Parameters \\\hline
    U-Net & 57.582 & 0.998 & 29.598 & 0.874 & 10.401M \\
    L-AE  & 57.66 & 0.998 & 29.609 & 0.874 & 1.675M \\\hline\hline
    \end{tabular}%
  \label{table3}%
\end{table}%

As we can see from Table \ref{table3}, whether in the Radon domain or in the image domain, L-AE has competitive performance compared with U-Net. Moreover, it significantly reduces model parameters, making it suitable for situations where computational resources are extremely limited. This exhibits our model’s ability in efficiently restoring CT images, thus adapting to the social trend of deploying portable medical devices.

\subsubsection{The Discriminator}
We establish discriminators in both two stages, hoping to further improve our model’s performance in restoring sparse-view CT data through the adversarial learning between the autoencoders and the discriminators. In order to verify this point of view, we send the test set into stage one where there is merely an autoencoder and compare its restoration results with ours, which can be seen from Table \ref{table4}. The images mentioned in Table \ref{table4} are reconstructed from the Radon data restored through stage one.

\begin{table}[htbp]
  \centering
  \caption{The Role of the Discriminator}
  \renewcommand{\arraystretch}{1.3}
    \begin{tabular}{m{5.2em}<{\centering}cccc}\hline\hline
    \multicolumn{1}{c}{} & \multicolumn{1}{m{3em}<{\centering}}{Radon PSNR} & \multicolumn{1}{m{3em}<{\centering}}{Radon SSIM} & \multicolumn{1}{m{3em}<{\centering}}{Image PSNR} & \multicolumn{1}{m{3em}<{\centering}}{Image SSIM} \\\hline
    L-AE Only & 48.904 & 0.985 & 28.448 & 0.871 \\
     \textbf{L-AAE} &  \textbf{57.660} &  \textbf{0.998} &  \textbf{29.609} &  \textbf{0.874} \\ \hline\hline
    \end{tabular}%
  \label{table4}%
\end{table}%

From the above table, we can realize the significance of our proposed discriminator, it indeed assists our model to achieve a better level of restoration under the evaluation of PSNR and SSIM. Its precise structure (refers to Sec II) also ensures a high degree of mobility, which enables our overall structure to be portable and accurate at the same time.

\subsubsection{The Two-step Architecture -- LS-AAE}
As we state in Sec II, this sort of cascaded two-step structure inherently embeds the motion of objects which can largely help to remove image artifacts due to the strong redundancy between these consecutive images. Consequently, we design an experiment with reference to \cite{tassano2020fastdvdnet} to prove this view. In the second stage, instead of sending five consecutive images into this two-step LS-AAE, we directly input them into a single Inpainting Block (SIB) that is slightly modified in the three-dimensional convolution part to handle five images, that means we adopt a stride of 2 in the Conv1\_1 layer (refers to Table \ref{table1}). The experimental results can be seen from Table \ref{table5} below.

\begin{table}[htbp]
  \centering
  \caption{Restoration Results of SIB and LS-AAE}
  \renewcommand{\arraystretch}{1.3}
    \begin{tabular}{p{4.5em}<{\centering}cc}\hline\hline
    \multicolumn{1}{c}{} & \multicolumn{1}{p{6em}<{\centering}}{Image PSNR} & \multicolumn{1}{p{6em}<{\centering}}{Image SSIM} \\\hline
    SIB   & 38.972 & 0.941 \\
     \textbf{LS-AAE} &  \textbf{40.305} &  \textbf{0.948} \\\hline\hline
    \end{tabular}%
  \label{table5}%
\end{table}%

Now the SIB no longer owns this built-in cascade structure to implicitly conduct motion estimation, it suffers from a obvious drop in PSNR and SSIM. Therefore, we can arrive at the conclusion that, LS-AAE manages to effectively improve model's capability of restoring CT images with its cascaded two-step architecture that inherently capture the motion between consecutive images.

\subsubsection{The 3D convolution in LS-AAE}
We mention in Sec II that, the extension of search regions from two-dimensional image neighborhoods to three-dimensional spatial neighborhoods provide extra information for image restoration. Also, extracting spatial features is conducive to remove streaking artifacts as the residual error in each image is correlated. In order to realize this extension of search regions, three-dimensional convolution is employed in every Inpainting Block of LS-AAE. To verify the cruciality of these three-dimensional convolutions, we conduct an experiment in which 3D convolution are replaced back to 2D convolution, where the number of input images is regarded as the number of input channel (refers to Table \ref{table2}). The inpainting results of these two models are shown in Table \ref{table6}.

\begin{table}[htbp]
  \centering
  \caption{Restoration Results of 2D and 3D LS-AAE}
  \renewcommand{\arraystretch}{1.3}
    \begin{tabular}{p{6em}<{\centering}cc}\hline\hline
    \multicolumn{1}{c}{} & \multicolumn{1}{p{6em}<{\centering}}{Image PSNR} & \multicolumn{1}{p{6em}<{\centering}}{Image SSIM} \\\hline
    2D LS-AAE & 39.472 & 0.944 \\
     \textbf{3D LS-AAE} &  \textbf{40.305} &  \textbf{0.948} \\\hline\hline
    \end{tabular}%
  \label{table6}%
\end{table}%

We can see that the inpainting outcome suffers from a drop about 0.9dB in PSNR, proving that three-dimensional convolutions assist model in restoring CT images to a certain extent without significantly consuming computational resources.

\subsubsection{The Image Interval of LS-AAE's Input}
In all the experiments above, we set the image interval between input consecutive CT images of LS-AAE to the default value of 1. However, we cannot help but wonder that whether increasing the interval value can help the model obtain more spatial information, thereby enhancing its ability in removing image artifacts. In the following experiment, we set this image interval $T$ to 1, 2, 3, 4 and 5 respectively, their corresponding results are shown in Table \ref{table7}.

\begin{table}[htbp]
  \centering
  \caption{The Image Interval's Effect on Restoration Results}
   \renewcommand{\arraystretch}{1.3}
    \begin{tabular}{p{4em}<{\centering}cc}
    \hline\hline
    \multicolumn{1}{c}{} & \multicolumn{1}{p{6em}}{Image PSNR} & \multicolumn{1}{p{6em}}{Image SSIM} \\ \hline
    $\bm{T=1}$   & \textbf{40.305} & \textbf{0.948} \\
    $T=2$   & 39.961 & 0.948 \\
    $T=3$   & 40.032 & 0.948 \\
    $T=4$   & 40.147 & 0.948 \\
    $T=5$   & 40.195 & 0.950 \\\hline\hline
    \end{tabular}%
  \label{table7}%
\end{table}%

It can be learnt from Table \ref{table7} that this hyperparameter T does not have much impact on the final restoration result. Spatial correlation seems to be well utilized when the image interval is set to 1, which would be a decent default choice.

\subsubsection{The Radon Domain VS. the Image Domain}
\begin{figure}[!htb]
\centerline{\includegraphics[width=\linewidth,scale=0.8]{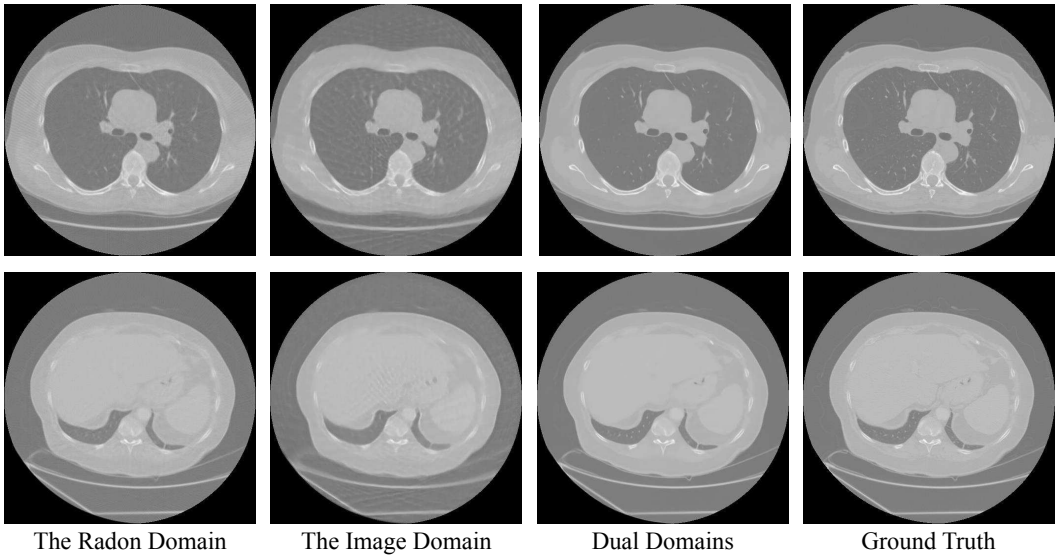}}
\caption{The intuitive restoration results obtained by different domains.}
\label{fig6}
\vspace{5pt}
\end{figure}

In this paper, we adopt a two-stage structure that combines the Radon domain and the image domain to obtain high-quality sparse-view CT images. Since each stage of the overall structure conducts restoration in their separate domains and both remarkably upgrade the restoration results, this leads us to think, what role do these two domains play? Subsequently, we feed our test set into these three structures: L-AAE in stage one that concentrates on the Radon domain, LS-AAE in stage two that focus on the image domain and of course, our overall structure that contains these two stages. The quantitative inpainting results of the above three structures can be referred from Table \ref{table8}, the intuitive outcome can also be seen in Figure \ref{fig6}.

\begin{table}[htbp]
\vspace{5pt}
  \centering
  \caption{Restoration Results obtained by different domains}
   \renewcommand{\arraystretch}{1.3}
    \begin{tabular}{p{9em}<{\centering}cc} \hline\hline
    \multicolumn{1}{c<{\centering}}{} & \multicolumn{1}{p{6em}<{\centering}}{Image PSNR} & \multicolumn{1}{p{6em}<{\centering}}{Image SSIM} \\ \hline
    The Radon Domain & 30.310 & 0.905 \\
    The Image Domain & 34.135 & 0.888 \\
    \textbf{Dual Domains} &  \textbf{40.305} &  \textbf{0.948} \\\hline\hline
    \end{tabular}%
  \label{table8}%
\end{table}%

It can be seen from above that, restoration in each domain has its pros and cons. For the Radon domain, it demonstrates its superiority in enhancing the structural similarity of images so as to perform well under the evaluation of SSIM. While as for the image domain, it exhibits great ability in alleviating distortion, thus has a relatively good performance under the evaluation of PSNR. Naturally, we acquire extraordinary restoration results when combining these two domains to merge their respective advantages. Besides, we solely utilize the spatial correlation in the Image domain due to our discovery that, the spatial information between continuous Radon slices has little impact on the final inpainting outcome. We suppose this is because the texture in Radon slices does not have much similarity with CT images, thus cannot be restored in this way.

\subsection{Methods Comparison}
\begin{figure}[!htb]
\centerline{\includegraphics[width=\linewidth,scale=0.8]{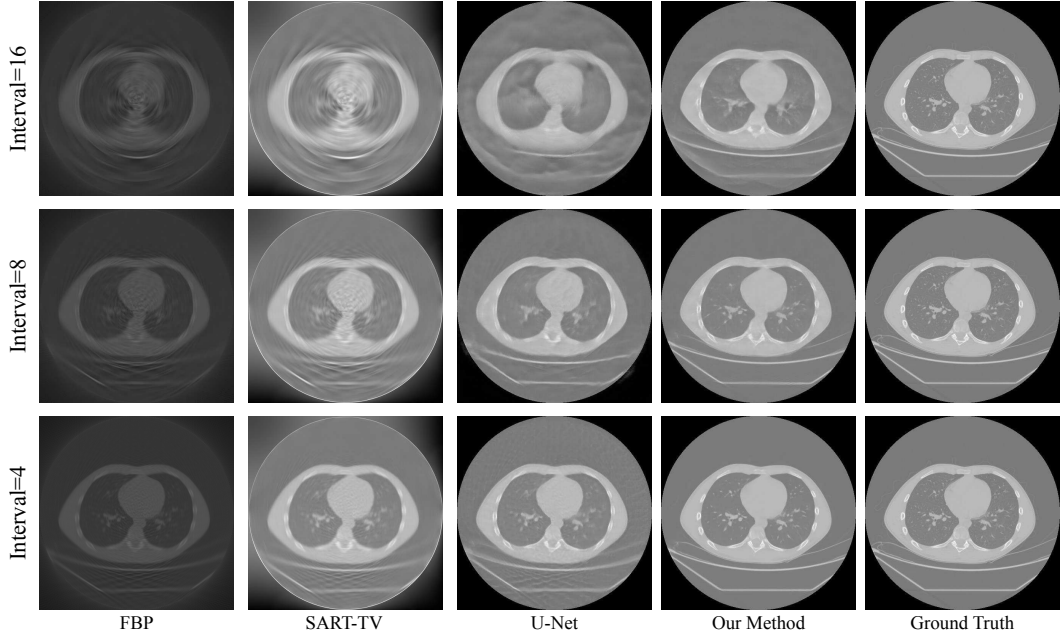}}
\caption{The intuitive restoration results of various methods at different sampling intervals.}
\label{fig7}
\end{figure}

After verifying the rationality of our overall structural design, we want to testify its robustness through applying it to sparse-view CT data with a higher level of sparsity, which means, conducting sparse sampling at intervals of 4, 8 and even 16 (the corresponding full-view Radon data has 180 projections). In addition, we compare our method with other current ones to prove its prominent capability of restoring sparse-view CT images and removing streaking artifacts. The experimental results are shown in Table \ref{table9}, and the intuitive outcome can be seen from Figure \ref{fig7}.

\begin{table}[htbp]
  \centering
  \caption{Methods Comparison}
  \renewcommand{\arraystretch}{1.3}
    \begin{tabular}{p{4.1em}<{\centering}p{2.7em}<{\centering}p{2.7em}<{\centering}p{2.7em}<{\centering}p{2.7em}<{\centering}p{2.7em}<{\centering}p{2.7em}<{\centering}} \hline\hline
    \multicolumn{1}{c}{\multirow{2}[4]{*}{}} & \multicolumn{2}{c}{Interval=4} & \multicolumn{2}{c}{Interval=8} & \multicolumn{2}{c}{Interval=16} \\
\cline{2-7}    \multicolumn{1}{c}{} & \multicolumn{1}{c}{PSNR} & \multicolumn{1}{c}{SSIM} & \multicolumn{1}{c}{PSNR} & \multicolumn{1}{c}{SSIM} & \multicolumn{1}{c}{PSNR} & \multicolumn{1}{c}{SSIM} \\
    \hline
    FBP   & 12.080 & 0.498 & 12.065 & 0.485 & 12.032 & 0.471 \\
    SART-TV & 19.179 & 0.665 & 19.061 & 0.632 & 18.777 & 0.602\\
    U-Net & 34.018 & 0.885 & 31.944 & 0.843 & 28.767 & 0.798 \\
    \textbf{Ours} & \textbf{40.305} & \textbf{0.948} & \textbf{37.633} & \textbf{0.937} & \textbf{34.052} & \textbf{0.910} \\ \hline\hline
    \end{tabular}%
  \label{table9}%
\end{table}%


As we can see, our method exhibits extraordinary capability of restoring sparse-view CT imaging, effectively removes streaking artifacts and outruns other methods by a large margin. Also, it can be applied to extreme sparsity while still obtaining prominent inpainting outcome. Particularly, our method still exceeds others when the sampling rate is one-fourth of them, thereby demonstrating its remarkable robustness and superiority.

\section{Conclusion}
In this paper, we propose a lightweight structure that efficaciously restores sparse-view CT with its two-stage architecture combining the Radon domain and the image domain. Most importantly, we groundbreakingly exploit the abundant spatial information existing between consecutive CT images, so as to achieve a remarkable restoration outcome even if our method encounters extreme sparsity.

In the first stage, a mobile model named L-AAE is proposed to complement the original sparse-view CT in the Radon domain, it adopts the inverted residual with linear bottleneck in order to significantly reduce computational resource requirements while maintaining outstanding performance. In the second stage, after reconstructing the restored full-view Radon data into images through FBP, we establish a lightweight model called LS-AAE. It is designed to implicitly conduct motion estimation and dig into the three-dimensional spatial neighborhood with a relatively low memory cost. Therefore, it manages to concentrates on fully utilizing the strong spatial correlation between continuous CT images, so as to productively remove streaking artifacts and finally acquire high-quality restoration results.

Eventually, for the sparse-view CT with a sampling interval of 4, we achieve a PSNR of 40.305 and a SSIM of 0.948, realizing a remarkable restoration result that can effectively eliminate image artifacts. In addition, our method also performs well when it comes to extreme sparsity (the sampling interval is 8 or even 16), exhibiting its prominent robustness.

\bibliographystyle{unsrt}
\bibliography{ref}

\end{document}